\documentclass[11pt]{article}
\usepackage[pdftex]{graphicx}
\usepackage{hyperref}
\usepackage{cite}
\usepackage{amsmath,amsfonts,amssymb,amsthm,nccmath,latexsym,mathtools}
\usepackage{xcolor}
\usepackage{amsmath}
\usepackage{cancel} 
\usepackage{ amssymb }
\usepackage{upgreek}
\usepackage{footnote}

\def\d{\delta}

\def\p{\partial}

\def\na{\nabla}
\def\T{\Theta}

\def\t{\tilde}

\def\be{\begin{equation}}
\def\ee{\end{equation}}
\def\ba{\begin{align}}
\def\ea{\end{align}}

\def\etc{{\it etc.}}
\def\ie{{\it i.e.~}}
\def\mT{$\mathcal{T}~$}
\begin{document}
\title{{\bf{\Large Thermogeometric study of van der Waals like phase transition in black holes: an alternative approach}}}
\author{
{\bf{\normalsize Krishnakanta Bhattacharya}}\footnote {\color{blue} krishnakanta@iitg.ac.in}, \ and \
{\bf{\normalsize Bibhas Ranjan Majhi}}\footnote {\color{blue} bibhas.majhi@iitg.ac.in}\\
Department of Physics, Indian Institute of Technology Guwahati, Guwahati 781039, Assam, India
}
\date{\today}
\maketitle
\begin{abstract}
It is well-known that the non-extremal anti-de-Sitter (AdS) black holes show various van der Waals type criticalities, such as the $P-V$, $T-S$, $Y-X$, $Q^2-\Psi$ \etc~In these cases, the phase space diagram of the relevant quantities are similar in behaviour to the isotherms on $P-V$ diagram for the usual van der Waals gas system. The existing thermogeometric descriptions of these criticalities for the black holes deal with the whole thermogeometric manifold and its curvature. However, while observing the criticality, one only notices the behaviour of the relevant phase space variables and keep all other thermodynamic quantities as fixed. Therefore it is quite natural to investigate the geometrical behaviour of the induced metric, defined by these constant macroscopic variables, corresponding to the whole manifold of the thermogeometry. Using geometrothermodynamics (GTD), we precisely address this issue. It is observed that the extrinsic curvature of this induced metric also diverges at the critical point. This provides an alternative but important aspect of the thermogeometric description of phase transition of black holes. The entire formalism in this paper is very general as it is valid for any arbitrary black hole which shows the van der Waals type phase transition.
\end{abstract}

\section{Introduction and Motivation}
The thermodynamic structure of black holes have been found decades ago. With proper identification of the entropy \cite{Bekenstein:1973ur}, temperature \cite{Hawking:1974sw, Bardeen:1973gs} and all other thermodynamic parameters, it can be shown that black hole shows several thermodynamic features that of the ordinary thermodynamics. Black hole phase transition is one such phenomenon which is being studied for long and still many interesting things are coming up with the due course of time. It was first Davies who claim that the phase transition occurs in black hole thermodynamics and he identified the critical point of the phase transition where the heat capacity shows an infinite discontinuity \cite{Davies:1989ey} (also see \cite{Banerjee:2011cz, Banerjee:2012zm, Majhi:2012fz, Lala:2012jp, Ma:2014tka, Azreg-Ainou:2014gja, Liu:2013koa, Mandal:2016anc}). There is another line of studying the black hole phase transition. In the AdS space, when the cosmological constant is regarded as the thermodynamic pressure \cite{Kastor:2009wy, Dolan:2010ha, Dolan:2011xt, Dolan:2011jm, Dolan:2012jh}, it was found that the $P-V$ diagram of black holes in the extended phase space looks exactly similar to that of the ordinary thermodynamics \cite{Kubiznak:2012wp, Kubiznak:2016qmn}, which implies a van der Waals phase transition in the black hole thermodynamics. Apart from the $P-V$ criticality, there are also a few other criticalities in black hole space which also are of van der Waals type, like  $T-S$ criticality \cite{Shen:2005nu, Zhang:2015ova, Zeng:2015tfj, Zeng:2015wtt, Spallucci:2013osa, Mo:2016apo, Kuang:2016caz}, $Q-\Phi$ criticality \cite{Ma:2016aat}, $Q^2-\Psi$ criticality \cite{Dehyadegari:2016nkd, Dehyadegari:2018pkb} \etc~ In addition, the van der Waals type phase transition has also been observed in \cite{Zhang:2014uoa}, where the phase transition of black holes has been studied in $AdS_5\times S_5$ spacetime by considering the cosmological constant as the number of colors in the boundary gauge theory and its conjugate quantity as corresponding chemical potential. Moreover, it has been found recently that the van der Waals phase transition can also be obtained in the black hole spacetime with a boundary which is not necessarily asymptotically AdS \cite{Pedraza:2018eey, Kastor:2018cqc}. Recently, the $P-V$ criticality has also been studied for black hole geometry with hyperscaling violation \cite{Pourhassan:2017bgs}. On a different note, we mention that the role of correction terms of the thermodynamic parameters in phase transition and in thermodynamic stability have been investigated in several spacetimes, such as: higher dimensional AdS black holes \cite{Upadhyay:2017vgk}, charged rotating AdS black holes \cite{Upadhyay:2018gee}, charged black holes in gravity's rainbow \cite{Upadhyay:2018vfu}, Ho\v{r}ava-Lifshitz black holes \cite{Pourhassan:2017qhq}, charged quasitopological and charged rotating quasitopological black holes \cite{Upadhyay:2017qmv}, non-rotating BTZ black holes \cite{Nadeem-ul-islam:2018mpc}, charged rotating BTZ black holes \cite{Ganai:2019lgc}, Schwarzschild-Beltrami-de Sitter black holes \cite{Sudhaker} \etc~

Both Davies type and van der Waals type phase transition has been widely explored under the light of the thermogeometry. The prescriptions of geometry was earlier incorporated in thermodynamics by Gibbs \cite{Gibbs}, Carath$\acute{\textrm{e}}$odory \cite{Carth}, Fisher \cite{Fisher} and Rao \cite{Rao}. Later Weinhold \cite{WEINHOLD} and Ruppeiner \cite{RUPP, Ruppeiner:1995zz} introduced Riemannian metrics in the thermodynamic phase space to study the black hole phase transition. The Weinhold metric was formulated as the Hessian of the internal energy and the Ruppeiner one is defined as the negative of the Hessian of the entropy. In fact, the Ruppeiner metric is a conformal one to the Weinhold metric with the conformal factor as the inverse temperature. Since in gravity the gravitational interaction results in the curvature in the spacetime, the same motivation is utilized while formulating thermogeometry \ie the thermodynamic interaction is described here by the curvature of the thermodynamic phase space and the critical points are identified as the point where the Ricci-scalar diverges \cite{Cai:1998ep, Aman:2003ug, Sarkar:2006tg, Myung:2008dma}. It was found later that the Weinhold metric and the Ruppeiner metric are not consistent with each other in some cases. For example \cite{Aman:2003ug}, in the case of the Kerr black hole, the Weinhold metric is flat and the Ruppeiner metric gives the divergence of the Ricci-scalar only at the extremal limit. Both of these metrics does not imply a phase transition as predicted by Davies where the heat capacity diverges. Moreover, recently in a general formulation, we have shown that the Weinhold metric cannot predict the extremal phase transition of black holes, whereas the Ruppeiner metric successfully can predict that \cite{Bhattacharya:2019awq}. This discrepancy in this two formulation is believed to appear due to that fact that both the Weinhold and the Ruppeiner metrics are not formulated in a Legendre-invariant way \cite{Quevedo:2006xk, Quevedo:2007mj, Quevedo:2008xn, Quevedo:2008ry, Alvarez:2008wa, Quevedo:2011np, Quevedo:2017tgz}. Since the thermodynamics is invariant due to the Legendre transformation, Quevedo {\it et al.} suggested that Legendre-invariance must be maintained while formulating a thermo-geometric metric and provided the prescription to formulate it in a Legendre-invariant way. Later Quevedo's formulation (popularly known as the geometrothermodynamics (GTD)) was widely accepted and it was shown that at the critical point, the Ricci scalar of the Legendre-invariant metric diverges \cite{Zhang:2015ova,Mo:2016apo, Shen:2005nu, Quevedo:2016cge, Quevedo:2016swn, Banerjee:2016nse}. Recently, accounting the corrections of the thermodynamic parameters, the correction in geometrothermodynamics has been shown for $f(R)$ gravity \cite{Upadhyay:2018bqy}. In one of our earlier works \cite{Bhattacharya:2017hfj}, we have constructed the Legendre invariant theromogeometrical manifolds for the van der Waals type phase transitions in AdS black holes. It was shown that the corresponding Ricci scalar diverges at the critical point. The whole analysis does not take into account the explicit form of the black hole metric. In addition to Weinhold's, Ruppeiner's and Quevedo's formulation of thermogeometrical metric, there is another popular formulation of thermogeometrical metric proposed by Hendi and collaboration \cite{Hendi:2015eca, Hendi:2015rja}, which is known as HPEM (Hendi-Panahiyan-Eslam Panah-Momennia) metric. Recently, the thermogeometric description using HPEM metric has been provided for the black holes in Rastall gravity \cite{Soroushfar:2019ihn} and also for the BTZ black holes in the context of massive gravity’s rainbow \cite{Hendi:2016hbe}. 

The present work is motivated from the following ideas. Recall that to observe the van der Waals type criticality, most of the thermodynamic quantities are kept fixed to a certain value and the determination of the critical point actually depends on a fewer thermodynamic parameters. Let us give a more specific example to clarify our argument. Consider the $P-V$ criticality of an arbitrary AdS black hole. The complete thermogeometric manifold of this black hole will consist of pressure $P$, volume $V$, entropy $S$, temperature $T$, multiple charges (and angular momentum) $Y_i$'s, corresponding potentials (and angular velocity) $X_i$'s \etc~But, while observing the criticality we fix all $X_i$'s and $Y_i$'s to a particular value and plot the isotherms on the $P-V$ plane. The critical point is confirmed on the critical isotherm $T=T_c$ from the set of two conditions $(\p P/\p V)_{T_c}=0=(\p^2 P/\p V^2)_{T_c}$. Therefore, we see that the several coordinates in the thermogeometric phase space does not play any role to determine the critical point.  Now, it will be interesting to see what happens if we fix those coordinates in the thermogeometric manifold to form a hypersurface (or line) and study the behaviour of the extrinsic curvature of that particular hypersurface (or line). In this regard let us mention that, recently the property of the extrinsic curvature (of a particular surface) has been observed for Reissner-Nordstr$\ddot{\textrm{o}}$m AdS black hole \cite{Mansoori:2016jer} in the case of Davis type phase transition, where it has been found that the extrinsic curvature of the Ruppeiner metric is divergent on the same points where the heat capacity diverges. Moreover, this result is further extended for any arbitrary black holes \cite{Zhang:2018djl}, where Davies' type criticality exists. Taking the same Legendre invariant metric which results in a divergent Ricci scalar at the Davies' critical point, it was shown \cite{Zhang:2018djl} that the extrinsic curvature of the same metric also diverges on the same critical point. Therefore, it is clear that one type of black hole criticality (Davies' type) can be described in terms of the extrinsic curvature of the thermodynamic phase space. 

Then the question, which naturally appears in one's mind is that whether the same analysis is true for the other criticality of black holes \ie for the van der Waals type of black hole criticality. In the present analysis, we found it is indeed possible to predict the black hole criticality in terms of the extrinsic curvature of the phase space. Therefore, our analysis further consolidates the previous claims \cite{Mansoori:2016jer, Zhang:2018djl} that the presence of black hole criticality is not only predicted by the divergence of the scalar curvature of the whole manifold, but also it can be predicted alternatively from the behaviour of the extrinsic curvature (of same thermo-geometric metric) on a particular hypersurface (the choice the hypersurface will be explained later in our analysis). Among several choices of defining thermogeometrical metric, we adopt the Legendre-invariant formulation as proposed by Quevedo \cite{Quevedo:2006xk, Quevedo:2007mj, Quevedo:2008xn, Quevedo:2008ry, Alvarez:2008wa, Quevedo:2011np, Quevedo:2017tgz}.

We have organized our paper in the following way. On the following section we provide the brief idea on how to formulate the Legendre-invariant metrics on the thermodynamic phase space. Also, there we mention the general form of the metric which we have used in our analysis. In the subsequent section, we study all the van der Waals type phase transition case by case: firstly the $P-V$ criticality and then $T-S$, $Y-X$ and $Q^2-\Psi$ criticalities follow. In each type criticality, we show that one can construct a pair of Legendre-invariant metrics (in fact, those metrics were obtained in our earlier work \cite{Bhattacharya:2017hfj} and we use those for the present analysis) and show that the critical point can be identified where the extrinsic curvatures of those metrics diverge simultaneously. Thereafter, we provide the conclusion of our work.

{\it Notations:} In our paper $P$ stands for pressure; $V$ for volume; $T$ for temperature; $S$ for entropy; $Y_i$ for multiple charges, angular momentum \etc~and  $X_i$, which is conjugate to $Y_i$, refers to the potentials due to those charges, angular velocity \etc
\section{Thermodynamic geometry in a Legendre invariant way: a brief review}

To formulate a thermo-geometric metric in a Legendre-invariant way, one is firstly needed to define a thermodynamic phase space \mT. The coordinates of \mT are defined as $Z^A=\Big(\Phi, \t X^a, \t P^a\Big)$~. Here $\Phi$ can be a thermodynamic potential or some other thermodynamic parameter, $\t X^a$ are the variables on which $\Phi$ depends and $\t P^a=\p \Phi/\p \t X^a$ are the conjugate variables. Now, the Legendre transformation in \mT is given by the following set of transformation
\begin{eqnarray}
&& (\Phi, \t X^a, \t P^a)\rightarrow (\Phi', \t X'^a, \t P'^a)
\\
&& \Phi=\Phi'-\delta_{ab}\t X'^a\t P'^b~, \t X^a=-\t P'^a~, \t P^a=\t X'^a~. \label{LEGT}
\end{eqnarray}
The canonical contact structure on the phase space \mT is defined by the fundamental Gibbs 1-form $\T=d\Phi-\sum_{a,b}\d_{ab}\t P^ad\t X^b$, which is Legendre invariant for the given transformation in \eqref{LEGT}. One can now define several Legendre-invariant thermo-geometric metric in \mT and the choice of those metrics are not unique \cite{Quevedo:2006xk}. Here we discuss a particular type of Legendre-invariant metric having the following structure
\begin{align}
G=\T^2+\Big(\lambda\sum_{a,b}\xi_{ab}\t P^a\t X^b\Big)\Big(\sum_{c,d}\eta_{cd} d\t P^cd\t X^d\Big)~. \label{GENMET1}
\end{align}
One can verify that the above metric \eqref{GENMET1} is invariant under the Legendre transformation \eqref{LEGT}. For the simplicity in the calculation we have taken $\lambda=1$, $\xi_{ab}=$diag$(1, 1, . . . . 1)$ and $\eta_{ab}=$diag$(-1, 1, ....1)$~. The space of equilibrium $\mathcal{E}$ is defined by the mapping $\upvarphi:\mathcal{E}\longrightarrow$ \mT with the constraint of the thermodynamic relation $d\phi=\sum_i\t P^id\t X^i$. Then further expanding $d\t P^i=\sum_j(d^2\Phi/d\t X^id\t X^j)d\t X^j$ one obtains the expression of the thermo-geometric metric in the equilibrium space as
\begin{align}
g=\upvarphi^{*}(G)=\Big(\sum_a \t X^a\frac{\p \Phi}{\p \t X^a}\Big)\Big(\sum_{c,d,e}\eta_{cd}\frac{d^2\Phi}{d\t X^cd\t X^e}d\t X^dd\t X^e\Big)~. \label{GENMET2}
\end{align}
Here, $\upvarphi^{*}$ is the pullback of $\upvarphi$~. Also, note that while obtaining the metric \eqref{GENMET2} we have explicitly used $\lambda=1$ and $\xi_{ab}=$diag$(1, 1, . . . . 1)$. 

 The above metric \eqref{GENMET2}, which has been defined in the equilibrium phase space, is the general form of the metric that we have used in our analysis. In the following section we explicitly present the form of thermo-geometric metric for different types of van der Waals criticalities (such as the $P-V$ criticality, $T-S$ criticality, $Y-X$ criticality, $Q^2-\Psi$ criticality \etc). Remember that these have been reported earlier in \cite{Dehyadegari:2018pkb, Bhattacharya:2017hfj} using the above prescription. So we shall just mention the metrics without any details. Below, we study the behaviour of the extrinsic curvature of the relevant hypersurface of these metrics near the critical point and show that it (extrinsic curvature) diverges on the critical point.

\section{Behaviour of the extrinsic curvature at the thermodynamic criticality of black holes}
In this section, we obtain the thermogeometric description of the van der Waals type criticality from the behaviour of the extrinsic curvature near the critical point. We have made this analysis for several van der Waals type criticalities that are observed in black hole spacetimes. Firstly we do it for the $P-V$ criticality and then the same technique is followed for other criticalities: $T-S$, $Y-X$, $Q^2-\Psi$ \etc


\subsection{$P-V$ criticality}
The presence of $P-V$ criticality in black hole spacetime was demanded earlier in \cite{Chamblin:1999tk, Chamblin:1999hg} where $P$ was identified as the inverse of the Hawking temperature (\ie $1/T_H$) and $V$ as the horizon radius (\ie $r_H$). This analysis mainly relies on the similarity of the graph of the $1/T_H$ and the graph of $P$ in van der Waals phase transition. Although the analysis apparently is intriguing, its main limitation was the lack of a proper canonical definition of $P$ and $V$ variables.
Later on, when the cosmological constant was taken as a variable and included in the extended version of first law \cite{Kastor:2009wy, Dolan:2010ha, Dolan:2011xt, Dolan:2011jm, Dolan:2012jh} (also see our work \cite{Bhattacharya:2017hfj}, where the first law in the extended phase space has been obtained from the conserved Noether current following Wald's method), the canonical definition of $P$ and $V$ was provided and was shown that the van der Waals $P-V$ criticality exists in black hole spacetime \cite{Kubiznak:2012wp, Kubiznak:2016qmn, Majhi:2016txt, Bhattacharya:2017nru} and at the critical point, two independent conditions coincide \ie $(\p P/\p V)_{T, Y_i}=0=(\p^2 P/\p V^2)_{T, Y_i}$~.

 In order to describe the $P-V$ criticality (of any arbitrary black hole in AdS spacetime) thermogeometrically, the proper Legendre-invariant thermo-geometric metric was obtained by us in \cite{Bhattacharya:2017hfj}, where $\Phi$ in \eqref{GENMET2} was taken as the Helmholtz free energy which is a function of volume $V$, temperature $T$ and the charges $Y_i$~. The metric is given as follows.
 \begin{align}
g^{(1)}_{ab}=(-PV-ST+\sum_iX_iY_i)[-F_{VV}dV^2+F_{TT}dT^2
 \nonumber
 \\
 +\sum_{i,j}F_{Y_iY_j}dY_idY_j+2\sum_iF_{TY_i}dTdY_i]~,
 \label{PVMETRIC1}
 \end{align}
 where $F$ is the Helmholtz free energy, $F_V=\Big(\frac{\p F}{\p V}\Big)_{T, Y_i}$, $F_{VV}=\Big(\frac{\p^2 F}{\p V^2}\Big)_{T, Y_i}$ and so on. When the criticality condition $P_V=(\partial P/\partial V)_{T, Y_i}=0$ is satisfied (\ie $F_{VV}=0$), we found \cite{Bhattacharya:2017hfj} that the Ricci-scalar $R_1^{(PV)}$ corresponding to the thermo-geometric metric \eqref{PVMETRIC1} diverges with the maximum order of divergence as $R_1^{(PV)}|_{max. diver.} \sim \mathcal{O}(1/F_{VV}^2)~.$ 
 
 The other criticality condition $P_{VV}=0$ is described by the other thermo-geometric metric, which is obtained by taking $\Phi$ as $P$ and the metric is given as follows.
 \begin{align}
g^{(2)}_{ab}
=(VP_V+TP_T+\sum_iY_iP_{Y_i})(-P_{VV}dV^2
\nonumber
\\
+P_{TT}dT^2+\sum_{j,k}P_{Y_jY_k}dY_jdY_k+2\sum_lP_{TY_l}dTdY_l)~.
\label{PVMETRIC2}
\end{align}
The scalar curvature of the metric \eqref{PVMETRIC2} diverges when $P_{VV}=0$ and the maximum order of divergence is given as $R_2^{(PV)}\sim\mathcal{O}(1/P_{VV}^2)$. Thus, according to our earlier analysis \cite{Bhattacharya:2017hfj}, the critical point (of $P-V$ criticality of black holes) in the manifold of thermodynamic phase space can be identified as the point where the Ricci-scalar corresponding to two different metrics \eqref{PVMETRIC1} and \eqref{PVMETRIC2} diverge simultaneously.

Let us now check whether the thermogeometric description can be given from the analysis of the extrinsic curvature, as was the case for the Davies' type criticality \cite{Mansoori:2016jer, Zhang:2018djl}. However, the method we follow here, is different from the previous works \cite{Mansoori:2016jer, Zhang:2018djl}. We shall fix those coordinates of the thermogeometric manifold, which are kept constant while the criticality conditions are obtained. This will imply a hypersurface (or line, in fact it turns out to be a line in the present case) being embedded in the total manifold. We shall calculate the extrinsic curvature of that line and show that the quantity diverges near the critical point. In the earlier cases \cite{Mansoori:2016jer, Zhang:2018djl}, the chosen surface are obtained in a different way. There is one more basic difference. Since, the van der Waals type criticality condition is comprised of two independent conditions, we need to analyze two extrinsic curvature of different thermo-geometric metrics (as we shall see later) to describe the criticality. Let us now proceed towards the main analysis.

Here, we consider the same metrics \eqref{PVMETRIC1} and \eqref{PVMETRIC2} like our previous analysis. We show that the alternative thermogeometric description (from the divergence of the extrinsic curvature) can also be obtained from the same thermo-geometric metrics, which is similar to \cite{Zhang:2018djl} (Davies' type criticality). Now, while observing the $P-V$ criticality of black holes, recall that we fix all the charges (and angular momentum) to a particular value (see FIG. 6 of \cite{Kubiznak:2012wp}) and see the behaviour of the isothermal curves on the $P-V$ plane. If we consider the whole phase space as a thermogeometric manifold described by \eqref{PVMETRIC1}, then the $P-V$ plane will be described by the by the thermo-geometric metric given as 
\begin{align}
g^{(1)}_{ab}=-f(V, T)dV^2+h(V, T)dT^2~, \label{PVMETRIC11}
\end{align}
with $f(V, T)=F_{VV}(-PV-TS+\sum_iX_iY_i)$ and $h(V, T)=F_{TT}(-PV-TS+\sum_iX_iY_i)$. Note that all $X_i$ and $Y_i$ are constant in the expression of $f(V, T)$ and $h(V, T)$. Also, $S$ and $P$ are the functions of $V$ and $T$. Therefore, the $P-V$ plane of the thermodynamics will correspond here in thermogeometry as the $V-T$ plane, with the coordinates being the independent variables $V$ and $T$. Now, we draw $T=$ const. isotherms in the $P-V$ diagram. This will correspond to $T=$const. lines in the thermogeometric surface \eqref{PVMETRIC11}. We calculate the extrinsic curvature of those lines and see whether the behaviour of the quantity can predict the criticality at $T=T_c$~.

The unit normal of a $T=$const. line on \eqref{PVMETRIC11} will be given as $n_i=(n_V, n_T)=(0, \sqrt{h(V, T)})$. Therefore, the extrinsic curvature will be given as $K^{(1)}_{(P,V)}=\na_in^i$. Finally one can obtain
\begin{align}
K^{(1)}_{(P,V)}=\frac{1}{2f\sqrt{h}}\frac{\p f}{\p T}~. \label{EXT11}
\end{align}
Note that $\p f/\p T\neq 0$ in general (for example, it is non-zero for Reissner-Nordstr$\ddot{\textrm{o}}$m AdS black holes). As a result, the extrinsic curvature corresponding to the critical isotherm will diverge due to the fact that $f(V, T)\sim F_{VV}=0$ when the first criticality condition is satisfied. Thus, we conclude that not only the Ricci-scalar of the whole manifold diverge, but the extrinsic curvature of the surface corresponding to the critical isotherm will also diverge. However, if in any case $\p f/\p T$ turns out to be zero, then $0/0$ case will appear in \eqref{EXT11}. That case has to be treated differently and in that case, our analysis might not be valid. This is true for the following instances as well.

Let us now look at the other metric \eqref{PVMETRIC2} and calculate the extrinsic curvature in the similar way. Again, following the earlier reason as earlier, one can say that the $P-V$ plane will be described by the metric
\begin{align}
g^{(2)}_{ab}=-k(V, T)dV^2+l(V, T)dT^2~, \label{PVMETRIC22}
\end{align}
where $k(V, T)=P_{VV}(VP_V+TP_T+\sum_iY_iP_{Y_i})$ and $l(V, T)=P_{TT}(VP_V+TP_T+\sum_iY_iP_{Y_i})$~. As was the earlier case, the isotherms in the $P-V$ plane will correspond to the $T=$const. line on \eqref{PVMETRIC22}. We obtain the expression of the extrinsic curvature of those lines and see its behaviour on the thermogeometric plane \eqref{PVMETRIC22}.

On a $T=$const. line one can define unit normal as $n_i=(n_V, n_T)=(0, \sqrt{l(V, T)})$ and finally, the extrinsic curvature will be given as 
\begin{align}
K^{(2)}_{(P,V)}=\frac{1}{2k\sqrt{l}}\frac{\p k}{\p T}~.
\end{align}
Again, when the second criticality condition is satisfied, the extrinsic curvature corresponding to the second metric will diverge as $k\sim P_{VV}=0$ is one of the criticality conditions.

Therefore, from our analysis we can say that the divergence of the extrinsic curvature can also predict the presence of a critical point of van der Waals type, as was the case for the other criticality (Davies' type) \cite{Mansoori:2016jer, Zhang:2018djl}. But, unlike the Davies' criticality, as there are two independent critical conditions in van der Waals type criticality, we have to analyze the behaviour of the two extrinsic curvatures and the critical point will be thermogeometrically defined where both of these diverge simultaneously.


\subsection{$T-S$ criticality}
Recently, there have been many works done on the $T-S$ criticality of black holes in the AdS spacetime \cite{Shen:2005nu, Zhang:2015ova, Zeng:2015tfj, Zeng:2015wtt, Spallucci:2013osa, Mo:2016apo, Kuang:2016caz} \footnote{The $T-S$ criticality has also been obtained in the case where $S$ corresponds to the entanglemaent entropy \cite{Caceres:2015vsa}} (also see \cite{Zeng:2016aly}, which have studied $T-S$ criticality for the black hole spacetime with higher derivative cuverture). This criticality is believed to appear because of the fact that the $P-V$ and the $T-S$ space are dual to each other \cite{Spallucci:2013osa}. However, while studying this criticality, the phase space is taken as the non-extended one \ie the cosmological constant is taken as the true constant in the theory. The $T-S$ criticality is also van der Waals type as the criticality conditions are implied by the similar conditions $(\p T/\p S)_{Y_i}=0=(\p^2 T/\p S^2)_{Y_i}$ ~. The $T-S$ criticality is described (as shown in \cite{Bhattacharya:2017hfj}) thermogeometrically by the set of two metrics. The first one is given as
\begin{align}
g^{'(1)}_{ab}=(SE_S+\sum_iY_iE_{Y_i})\Big(-E_{SS}dS^2+\sum_{i,j}E_{Y_iY_j}dY_idY_j\Big)~. \label{TSMETRIC1}
\end{align}
The above metric \eqref{TSMETRIC1} can be obtained from the general expression \eqref{GENMET2} when one identifies $\Phi$ as the internal energy $E$, which is a function of the entropy $S$ and the charges $Y_i$ in the non-extended phase space.
The another thermo-geometric metric is obtained from \eqref{GENMET2} identifying $\Phi$  as $T$. The metric is given as
\begin{align}
g^{'(2)}_{ab}=(ST_S+\sum_iY_iT_{Y_i})\Big(-T_{SS}dS^2+\sum_{j,k}T_{Y_jY_k}dY_jdY_k\Big)~. \label{TSMETRIC2}
\end{align}

 In this case, while observing the criticality in the $T-S$ plane (for example see Figure 1 of \cite{Kuang:2016caz}), we consider only one charge (say $Y$) while fixing the others. Therefore, in the thermogeometric manifold described by the metric $g^{'(1)}_{ab}$, the $T-S$ plane of thermodynamics will correspond to the thermogeometric $S-Y$ plane described by the metric
\begin{align}
g^{'(1)}_{ab}=-f'(S, Y)dS^2+h'(S, Y)dY^2~, \label{TSMETRIC3}
\end{align}
where, $f'(S, Y)=E_{SS}(ST+\sum_iX_iY_i)$ and $h'(S, Y)=E_{YY}(ST+\sum_iX_iY_i))$~. Note that the summation of charges includes all the charges which are kept constant on a particular value as well as the charge $Y$ which varies. Now, in the $T-S$ plane we draw the lines of $Y=$const. curves and the critical curve is described by $Y=Y_c$. Therefore, here we want to see the behaviour of the extrinsic curvatures corresponding to $Y=$const. curves of the thermogeometric plane \eqref{TSMETRIC3}.  The unit normal for $T=$const. lines will be given as $n_i=(n_S, n_Y)=(0, \sqrt{h'(S, Y)})$. Finally, the extrinsic curvature will be given as 
\begin{align}
K^{(1)}_{(T, S)}=\frac{1}{2f'\sqrt{h'}}\frac{\p f'}{\p Y}~.
\end{align}
Since $f'(S, Y)\sim E_{SS}=0$ on the critical point, the extremal curvature diverges there.

Let us now follow the same procedure for the thermogeometric manifold defined by the metric \eqref{TSMETRIC2}. In that case, the thermodynamic $T-S$ plane will be described by the metric (while all other charges except $Y$ are kept fixed),
\begin{align}
g^{'(2)}_{ab}=-k'(S, Y)dS^2+l'(S, Y)dY^2~, \label{TSMETRIC4}
\end{align}
with $k'(S, Y)=T_{SS}(ST_S+\sum_iY_iT_{Y_i})$ and $l'(S, Y)=T_{YY}(ST_S+\sum_iY_iT_{Y_i})$~. Again, we obtain the extrinsic curvature of $Y=$const. lines of the thermogeometric plane \eqref{TSMETRIC4}.

The unit normal for  a $T=$const. surface in the manifold described by \eqref{TSMETRIC4} is given as $n_i=(n_S, n_Y)=(0, \sqrt{l'(S, Y)})$ and the expression of the extrinsic curvature will be given as
\begin{align}
K^{(2)}_{(T, S)}=\frac{1}{2k'\sqrt{l'}}\frac{\p k'}{\p Y}~.
\end{align}
Again, one can show that when the second criticality condition $T_{SS}$ is satisfied, the extrinsic curvature $K^{(2)}_{(T, S)}$ diverges. Therefore, we can say that the $T-S$ criticality can be identified from the divergence of two extrinsic curvatures as well. Thus, our analysis for the $P-V$ criticality can be extended further for the $T-S$ criticality as well. Let us now look at another type of van der Waals type criticality, \ie the $Y-X$ criticality.

\subsection{$Y-X$ criticality}
The van der Waals type criticality has also been obtained for the charge-potential \cite{Ma:2016aat}  and also for angular momentum-angular velocity in the non extended phase space. We generally call it as the $Y-X$ criticality as, in our case, $Y_i$ represents all the charges, angular momentum \etc~and $X_i$ represents the potential, angular velocity \etc~The criticality conditions are given as $(\partial Y_i/\partial X_i)_{T, \bar X_i, \bar Y_i}=0=(\partial^2Y_i/\partial X^2_i)_{T, \bar X_i, \bar Y_i}$~. Here, $\bar Y_i$ implies the set of all the charges and angular momentum except $Y_i$ and similarly $\bar X_i$ implies the set of all angular velocity and potentials except $X_i$.
As we have obtianed in \cite{Bhattacharya:2017hfj}, the $Y-X$ criticality is described by a set of two metrics. The first one is given as
 \begin{align}
g^{''(1)}_{ab}=(-TS-\sum_kY_kX_k)\Big(-\sum_{ij}G_{X_iX_j}(T, X_i)dX_idX_j+G_{TT}(T, X_i)dT^2\Big)~. \label{YX1}
 \end{align}
 Here, $\Phi$ is identified as $G(T, X_i)$ -- the Gibbs free energy to obtain \eqref{YX1}.
 The other metric which describes the second criticality condition can be obtained by identifying $\Phi$ as one of the charges (say the m'th charge $Y_m$) and the metric is given as
\begin{align}
g^{''(2)}_{ab}=(\sum_kY{_m}_{X_k}X_k+Y{_m}_TT)\Big(-\sum_{i,j}Y{_m}_{X_iX_j}dX_idX_j+Y{_m}_{TT}dT^2\Big)~. \label{YX2}
\end{align}

As shown in the Figure 1 of \cite{Ma:2016aat}, in this case the criticality relation is obtained between a particular charge $Y$ and its potential $X$ while all other charges and potentials are kept fixed. Then in $Y-X$ plane we draw the isothermal lines and shows that on the critical isotherm the critical point is located at the point where the criticality conditions are satisfied. Therefore, the $Y-X$ plane will correspond to the thermogeometric plane which will be described by the metric 
\begin{align}
g^{''(1)}_{ab}=-f''(X, T)dX^2+h''(X, T)dT^2~, \label{YX3}
\end{align}
where $f''(X, T)=G_{XX}(-TS-\sum_kY_kX_k)$ and $h''(X, T)=G_{TT}(-TS-\sum_kY_kX_k)$~. Again, the isothermal lines in the thermodynamic $Y-X$ space will correspond to the $T=$const. lines in the thermogeometric surface described by the metric \eqref{YX3}. The expression of the extrinsic curvature on $T=$const. lines can be obtained following the earlier method. Its expression will be given as
\begin{align}
K^{(1)}_{(Y,X)}=\frac{1}{2f''\sqrt{h''}}\frac{\p f''}{\p T}~. \label{EXT31}
\end{align}
Since $f''(X, T)\sim G_{XX}$ vanishes on the critical point, the extrinsic curvature in \eqref{EXT31} will diverge on the critical point.

The $Y-X$ plane in the thermogeometric manifold \eqref{YX2} will be described by the metric (identifying $Y_m=Y$)
\begin{align}
g^{(2)}_{ab}=-k''(X, T)dX^2+l''(Y, T)dT^2~, \label{YX4}
\end{align}
where $k''(X, T)=Y_{X_iX_j}(\sum_kY_{X_k}X_k+Y_TT)$ and $l''(X, T)=Y_{TT}(\sum_kY_{X_k}X_k+Y_TT)$~. The extrinsic curvature of $T=$const. line will be obtained as
\begin{align}
K^{(2)}_{(Y, X)}=\frac{1}{2k''\sqrt{l''}}\frac{\p k''}{\p T}~.
\end{align}
Similar to the earlier case, the extrinsic curvature diverges on the critical point. Thus, we see that our analysis continue to be valid for the $Y-X$ criticality as well. However, apart from the charge-potential criticality, another $Q^2-\Psi$ criticality is obtained in black hole, which is of van der Waals type. In the following, we see whether our analysis fit for that criticality.

\subsection{$Q^2-\Psi$ criticality}
Recently in \cite{Dehyadegari:2016nkd, Dehyadegari:2018pkb}, it has been shown that for the Reissner-Nordstr$\ddot{\textrm{o}}$m AdS black holes, van der Waals type $Q^2-\Psi$ criticality exists when the cosmological constant are kept fixed. Here $Q$ is the charge of the black hole. In that case, the thermodynamic first law comes out as $dM=TdS+\Psi dq$ and the criticality conditions are given as $(\p q/\p \Psi)_{T_c}=0=(\p^2 q/\p \Psi^2)_{T_c}$~, where $q=Q^2$~. Thermogeometric description in terms of the divergent Ricci-scalar has been obtained in that work \cite{Dehyadegari:2018pkb}. The first thermo-geometric metric has been obtained identifying $\Phi$ as $K$~, where the thermodynamic potential $K$ is defined as $K=M-TS-q\psi$ and the metric is given as  (see eq. (34) of \cite{Dehyadegari:2018pkb})
\begin{align}
g_{ab}^{'''(1)}=-f'''(\Psi, T)d\Psi^2+h'''(\Psi, T)dT^2~, \label{Q2PSI1}
\end{align}
where $f'''(\Psi, T)=K_{\Psi\Psi}(\Psi K_{\Psi}+TK_T)$ and $h'''(\Psi, T)=K_{TT}(\Psi K_{\Psi}+TK_T)$~. Also, using the first law of black hole mechanics, one can obtain $(\p q/\p \Psi)_{T_c}=-K_{\Psi\Psi}$. As a result, $f'''(\Psi, T)$ vanishes on the critical point. Now, on $Q^2-\Psi$ plane we draw isotherms and show that the van der Waals type criticality exists for $T=T_c$~. Therefore, on the thermogeometric plane \eqref{Q2PSI1} we obtain the extrinsic curvature of $T=$const. curves, which corresponds to the isotherms on the thermodynamic $Q^2-\Psi$ plane. Following the earlier way, the expression of the extrinsic curvature can be obtained as
\begin{align}
K^{(1)}_{(Q^2,\Psi)}=\frac{1}{2f'''\sqrt{h'''}}\frac{\p f'''}{\p T}~. \label{EXT41}
\end{align}
Like the earlier cases, one can show that the extrinsic curvature diverges at the critical point as $f'''\sim K_{\Psi\Psi}= (\p q/\p \Psi)_{T_c}=0$ when the criticality conditions are satisfied~. 

The another criticality condition is thermogeometrically described by another metric when one takes $\Psi$ as $q$. In that case, the thermogeometric metric is obtained as (see eq. (38) of \cite{Dehyadegari:2018pkb})
\begin{align}
g_{ab}^{'''(2)}=-k'''(\Psi, T)d\Psi^2+l'''(\Psi, T)dT^2~, \label{Q2PSI2}
\end{align} 
where $k'''(\Psi, T)=q_{\Psi\Psi}(\Psi q_{\Psi}+Tq_T)$ and $l'''(\Psi, T)=q_{TT}(\Psi q_{\Psi}+Tq_T)$~. Also, note that $q_{\Psi\Psi}=(\p^2 q/\p \Psi^2)_{T_c}$, which implies $k'''(\Psi, T)$ vanishes on the critical point. Again we compute the extrinsic curvature of $T=$const. lines on the thermogeometric surface \eqref{Q2PSI2} which corresponds to the isothermal curves on the thermodynamic $Q^2-\Psi$ plane. It is obtained as
\begin{align}
K^{(2)}_{(Q^2,\Psi)}=\frac{1}{2k'''\sqrt{l'''}}\frac{\p k'''}{\p T}~.\label{EXT42}
\end{align}
One can show the extrinsic curvature of \eqref{EXT42} diverges when $k'''\sim (\p^2 q/\p \Psi^2)_{T_c}=0$ at the critical point. As a result, on the critical point both the extrinsic curvatures of \eqref{EXT41} and \eqref{EXT42} diverge simultaneously.

Thus, in this section, we have studied the behaviour of the extrinsic curvature for each van der Waals type criticality in the thermogeometric manifold. From the viewpoint of geometrothermodynamics, firstly we have studied the $P-V$ criticality. To describe both the critical conditions geometrically, we obtained a pair of thermogeometric surface which corresponds to the $P-V$ thermodynamic place, where different isotherms are drawn. Then the extrinsic curvatures of $T=$const. lines of the thermogeometric surface, which corresponds to various isotherms in the thermodynamic $P-V$ plane, are calculated. Thereafter, we show that both the extrinsic curvatures corresponding to the critical isotherm diverge at the critical point. Then we extend this analysis for other van der Waals like phase transition in black hole thermodynamics \ie $T-S$ criticality, $Y-X$ criticality and, finally, the $Q^2-\Psi$ criticality. In every case we show that the critical point corresponds to the simultaneous divergence two extrinsic curvature defined in two  different thermogeometric manifold. 

\section{Conclusion}
The presence of van der Waals type phase transition in black hole thermodynamics is a remarkable revelation by the physicists. Over the years, many works are going on in this area and many interesting information are getting revealed with the due course of time. Studying the various properties of black hole thermodynamics in terms of the thermogeometry is a recent trend. As we have mentioned earlier, the underlying idea of this formalism (thermogeometry) is to describe the thermodynamic interaction in terms of the curvature of the phase space. Both the Davies' type as well as the van der Waals type phase transition has been rigorously studied using the geometrothermodynamics, where it has been unanimously accepted that the critical point of the black hole phase transition corresponds to the divergence of the Ricci scalar of the thermodynamic phase space \cite{Zhang:2015ova,Mo:2016apo, Hendi:2015eca, Hendi:2015rja, Shen:2005nu, Quevedo:2016cge, Quevedo:2016swn, Banerjee:2016nse, Bhattacharya:2017hfj}. However, as shown in the recent analysis \cite{Mansoori:2016jer, Zhang:2018djl} for Davies' type criticality, we also have formulated an alternative thermogeometric description for the van der Waals type phase transition from the study of the extrinsic curvature tensor near the critical point. 

 As we have mentioned earlier, the major inspiration to explore in this direction is the following. We know that the criticality conditions for the van der Waals phase transition does not depend on the all of the thermodynamic quantities present in the theory as most of those are kept fixed to certain values. Eventually, the criticality conditions are obtained by varying a fewer thermodynamic quantities. Therefore, if the divergence of the curvature is a signature of the presence of a critical point, then it should be reflected not only on the curvature of the whole manifold (which is the present conception), but also on the curvature of the thermogeometric surfaces, which are obtained by fixing those thermodynamic parameters (which are kept fixed while determining the critical point). This was the major motivation of our work and we finally succeeded in obtaining the alternative thermogeometric description from the study of the extrinsic curvature instead of the curvature of the whole manifold.

 To provide the (alternative) thermogeometric description of van der Waals critical point, we had to formulate proper thermo-geometric metrics, which are convenient for the description of the critical point. In our case, the suitable Legendre-invariant metrics was obtained in our earlier work \cite{Bhattacharya:2017hfj}, where we have shown that the corresponding Ricci-scalar diverges at the critical point. Considering the thermogeometric manifold defined by those metrics, we have shown that if certain thermodynamic parameters are kept fixed, which is the case while obtaining the critical point, then it corresponds to a line in the thermogeometric manifold and the extrinsic curvature of the line diverge at the critical point. Since the van der Waals crticality is described by two independent conditions, we had to take the metrics of two different thermogeometric manifold and the extrinsic curvatures were shown to diverge simultaneously at the critical point. Thus, we see that the presence of a critical point in the thermodynamics can  be described not only by the divergence of the Ricci scalar but it can also be described alternatively by the divergence of the extrinsic curvature. Moreover, the present analysis is valid not only for a particular black hole, but also for any arbitrary black hole in AdS spacetime which shows the van der Waals type phase transition. 
 
 There is a very important point, needs to be mentioned at the end.  Coincidentally, both the Ricci-scalar and the relevant extrinsic curvature of the metrics, presented here, diverge at the critical point. Therefore, although we have a precise motivation for looking the nature of extrinsic curvatures (this has been stated earlier), it seems to be ``obvious'' that the extrinsic curvature also behaves in a same manner at the critical point as the Ricci scalar. In this regard, we want to mention that the divergence of the extrinsic curvature does not guarantee the divergence of the Ricci-scalar or vice versa. Those two are completely different quantities, and bears different meaning in differential geometry. There exists examples of manifolds in gravity which show this clear difference. The simpliest example is the Schwarzschild metric. In $(1+1)$ dimensions, the extrinsic curvature of $r$ = constant surface of the Schwarzschild metric, given by $K^{(2)}= f'/(2\sqrt{f})$ with $f=(1-2M/r)$, diverges at the horizon $r=2M$; whereas the Ricci scalar $R^{(2)}\sim f''\sim 4M/r^3$ does not, rather it is finite. In $(1+3)$ dimensional situation one has $K^{(4)} \sim (1/r^2)(2r\sqrt{f}+f'/(2\sqrt{f}))$, which clearly diverges at $r=2M$; but Ricci scalar $R^{(4)}$ vanishes. Since the intrinsic curvature and the extrinsic curvature are different, one has to treat them differently. Hence, for our thermogeometric metrics, although the Ricci scalars diverges at the critical point (known earlier) we are not sure about the feature of relevant extrinsic curvatures at the same critical point. Having proper motivation and importance to study these extrinsic curvatures, which we have stated earlier, one needs to investigate them separately. This is what we have done in the present paper. In this sense, the present analysis is very important and quite natural. Thus, on a final note, we believe that our work will play a crucial role in the study of black hole phase transition.



\end{document}